\documentclass[aps,prl,twocolumn,nofootinbib]{revtex4-1}
\usepackage{graphicx}
\usepackage{amssymb,amsmath,enumerate}
\usepackage{color}
\usepackage{bm}
\newcommand{\be}{\begin{eqnarray}}
\newcommand{\ee}{\end{eqnarray}}

\begin{document}

\title{Effect of friction on random adhesive loose packings of micron-sized particles}

\author{Wenwei Liu$^{1}$, Yuliang Jin$^{2}$, Shuiqing Li$^{1}$\footnote{lishuiqing@tsinghua.edu.cn}, Sheng Chen$^{1}$, Hern\'an A. Makse$^{3}$}

\affiliation{ $^1$Key Laboratory for Thermal Science and Power Engineering of Ministry of Education, Department of Thermal Engineering, Tsinghua University, Beijing 100084, China \\ 
$^2$Department of Physics, Osaka University, Japan\\ 
$^3$Levich Institute and Physics Department, City College of New York, New York 10031, USA }

\begin{abstract}

The effect of friction on random packings of micron-sized spheres is investigated by means of adhesive contact dynamics simulation and statistical ensemble theory. The structural properties of the adhesive packings with different friction coefficient $\mu_{\rm f}$ can be well described by an ensemble approach based on a coarse-grained volume function. A mechanical equilibrium analysis demonstrates that the packing structures become denser when $\mu_{\rm f}\leq 0.01$, because of the prominent rearrangements arising from the relative sliding motion between contact particles. We propose a modified isostatic condition to account for the low coordination numbers of frictional packings of adhesive spheres obtained in the simulation. Together with the equation of state derived from the statistical ensemble approach, theoretical predictions of the packing properties of adhesive frictional particles are obtained, which are in good agreement with simulations.

\end{abstract}

\date{\today}

\maketitle


Random packings of uniform spherical particles have been studied to model the microstructure and bulk properties of simple liquids, metallic glasses and colloidal crystals \cite{Bernal59,Parisi10} as well as frictional granular materials \cite{Coniglio04}. The presence of friction substantially expands the volume fraction from the random close packing (RCP) limit at $\phi_{RCP}\approx 0.64$ \cite{Bernal59}, which is associated with the {\it jamming transition} around the J point or the J line \cite{O'Hern03,Parisi10,Torquato10,Song08}, to a lower bound identified as the random loose packing (RLP) at $\phi_{RLP}\approx 0.55$ \cite{Song08,Onoda90,Ciamarra08,Jerkins08,Dong06,Farrell10}. Furthermore, friction also has a significant impact on the mechanical equilibrium of random packings. The minimal average coordination number, $Z$, required to obtain static packing in d-dimension is within the range $d+1\leq Z \leq 2d$ \cite{Silbert02}. It is reported that a frictional packing is isostatic when we consider interactions between asperities on contacting particles \cite{Papanikolaou13} or exclude the fully mobilized contacts at Coulomb threshold \cite{Shundyak07,Henkes10}.

For sufficiently small particles, most packings in nature are subject to, not only friction, but also adhesive forces. For instance van der Waals forces become non-negligible and generally dominate interactions between micron-sized particles smaller than $10\mu m$. In this case, adhesive forces could distinctly change the macroscopic structural properties \cite{Li11,Marshall14}. Despite the ubiquitous application of adhesive particle packings in various areas of engineering, biology, agriculture and physical sciences \cite{Marshall14,Dominik97,Blum04}, limited investigations have been made to systematically study these packings \cite{Valverde04,Martin08,Parteli14,Liu15,Chen16}. The greatest challenge arises from the multi-coupling of adhesive force, friction and other interactions in the short-range particle-particle interaction zone. In the absence of experiments, discrete element simulation has become an efficient and accurate method to study the packing problems of micron-sized particles \cite{Parteli14,Liu15,Chen16}. Based on the widely accepted adhesive discrete element method (DEM) \cite{Johnson71,Derjaguin75,Brilliantov96,Brilliantov07,Li11,Marshall14}, both macroscopic and microscopic properties of adhesive particle packing can be studied in detail.

Simulations and experiments have found that the volume fraction for adhesive micron-sized particles can go far below the RLP limit and decrease with smaller sizes \cite{Parteli14,Valverde04}. The lowest packing fractions were obtained within $0.15-0.23$ with particle size ranging from $1.5\mu m$ to $7.8\mu m$ \cite{Parteli14, Valverde04}. The average coordination number of adhesive packings for particles smaller than $10\mu m$ lay below the isostatic limit $Z=4$. Not only particle size but also work-of-adhesion and particle velocity influence the packing properties, of which the combined effect was characterized via a dimensionless adhesion parameter ($Ad$) \cite{Liu15}. It was concluded that both $\phi$ and $Z$ went below the RLP limit with $Ad>1$, and further approached an asymptotic {\it adhesive loose packing} (ALP) limit $\phi_{ALP}=1/2^3, Z_{ALP}=2$, which was also interpreted through a statistical ensemble approach at a mean-field level \cite{Liu15,Baule13,Baule14}. However, the friction effect on these very loose packings of adhesive particles has not been fully discussed. Little attention has been paid to the problem of how the adhesive packings change for arbitrary friction.

To address the above questions, we perform a systematic investigation of the friction effect on random packings of micrometer sized {\it soft-sphere, non-Brownian}, uniform adhesive particles via discrete element simulation. The DEM framework used in this work is specifically developed for adhesive particles \cite{Li11,Marshall14,Marshall09}, in which both the translational and rotational motions of each particle in the system are considered on the basis of Newton's second law (see details in the Supplementary Information). The adhesive contact forces $F_A$ include three terms, the normal adhesive contact force $F_{\rm ne}$, the normal damping force $F_{nd}$ with model parameters validated by classic particle-surface impact experiments \cite{Li11,Marshall14}, and the tangential force due to the sliding friction. A JKR (Johnson-Kendall-Roberts) model is employed to account for $F_{\rm ne}$ between the relatively compliant micro-particles \cite{Johnson71}. The dissipative friction, including the sliding, twisting and rolling terms in the presence of adhesion, are all approximated by a spring--dashpot--slider model with model parameters given in \cite{Li11,Heim99,Sumer08}.

The adhesive DEM simulation starts with the successive random injection of 1,000 spheres of radius $r_p$ from a surface at a height $H=80r_p$ with an initial velocity $U_0$ under gravity. Periodic boundary conditions are set along the two horizontal directions of length $L=20r_p$. The particle size ranges from $1\mu m$ to $50\mu m$ and the work-of-adhesion $w$ is $0.1\sim 30mJ/m^2$. It should be noted that a low initial velocity of $U_0=0.5m/s$ is applied to guarantee a less considerable compaction resulted from particle inertia. The friction coefficient is changed from $\mu_{\rm f}=10^{-5}$ to $\mu_{\rm f}=10$ with other parameters fixed in each case.



\begin{figure}
\begin{center}
\includegraphics[width=9cm]{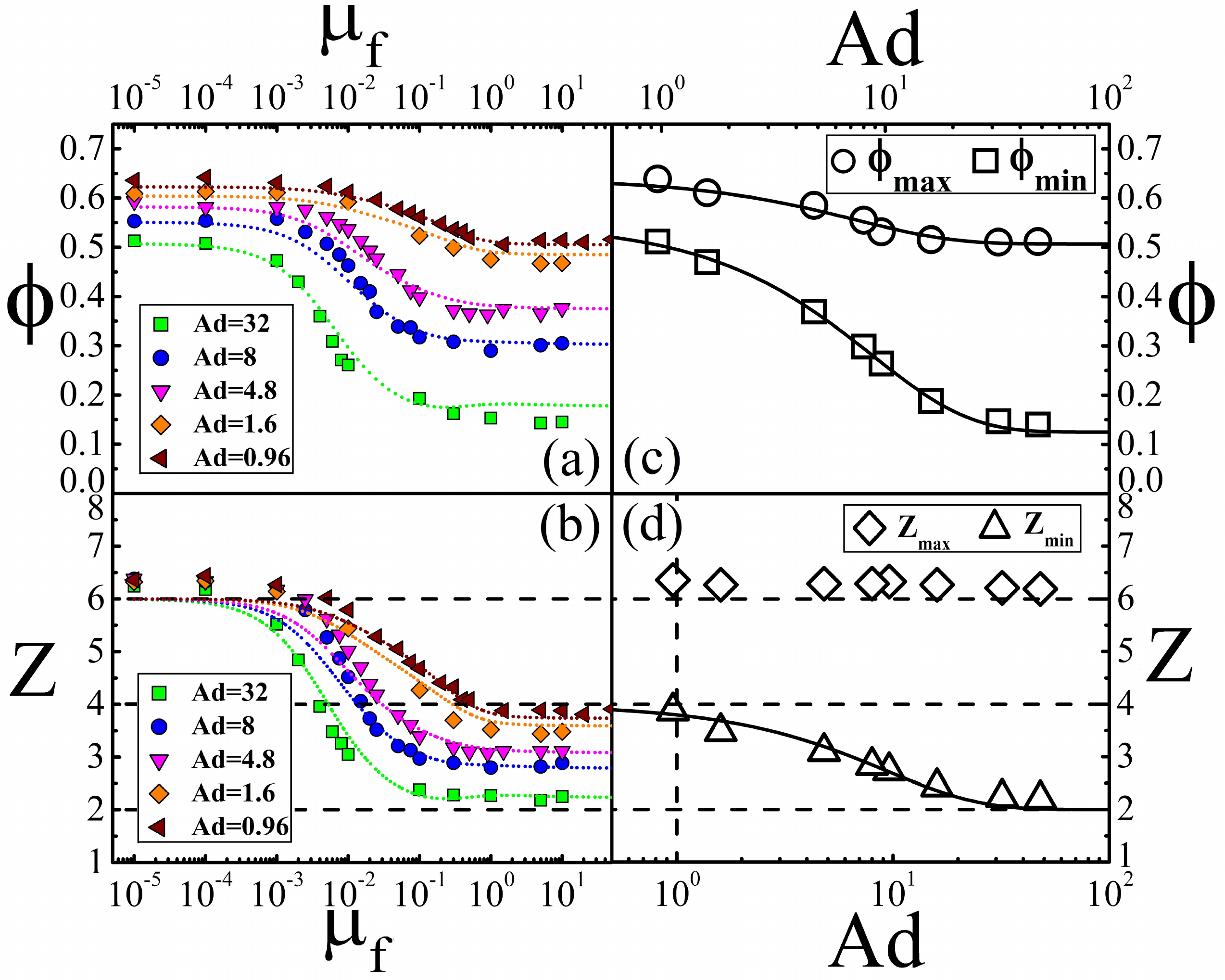}
\caption{\label{Fig:phiZ} Volume fraction $\phi$ (a) and average coordination number $Z$ (b) as a function of $\mu_{\rm f}$ with different $Ad$. The dotted lines are theoretical fitting lines. (c) and (d) show the variation of $\phi_{min}$, $\phi_{max}$ and $Z_{min}$, $Z_{max}$ with $Ad$, respectively.}
\end{center}
\end{figure}

The major result of this work is shown in Fig.\ref{Fig:phiZ}, which exhibits the volume fraction ($\phi$) and average coordination number ($Z$), with friction coefficient varying in the range $\mu_{\rm f}=10^{-5}\sim 10$. Here the dimensionless adhesion parameter $Ad=w/(2\rho_p U_0^2 R)$ is applied to quantify the combined effects of particle velocity, size and adhesion \cite{Li07, Liu15}, where $\rho_p$ is the mass density and $R$ is the reduced radius. It can be seen that with fixed $Ad$, both $\phi$ and $Z$ reach an upper limit when $\mu_{\rm f}\leq 10^{-4}$ and decrease with the increasing of $\mu_{\rm f}$ until a lower limit is obtained when $\mu_{\rm f}>1$ (panels (a) and (b)). The different upper and lower limit values of $\phi$ depend on the values of $Ad$ while only the lower limit values of $Z$ have similar dependence on $Ad$ (panels (c) and (d)). The upper limits of $Z$ collapse around $Z=6$, which corresponds to the isostatic limit of frictionless spheres packing. This indicates that adhesion has no effect on the coordination number when friction is small. Furthermore, all of the limit values $\phi_{min}$, $\phi_{max}$ and $Z_{min}$ are found to follow the same exponential law (solid lines in panels (c) and (d)), i.e. 
\be
\chi_m=\chi_{0,m}+\chi_{1,m}exp(-\lambda Ad), 
\label{eq:phiZmaxmin}
\ee
differing in the fitting parameters $\chi_{0,m}$, $\chi_{1,m}$ and $\lambda$, where $\chi$ represents either $\phi$ or $Z$ and the subscript $m$ means maximum or minimum (see details in SI). It is also interesting that the packing properties fall between RLP and RCP when $Ad=0.96$, confirming the critical value of $Ad=1$ that distinguishes a unique adhesion controlled packing regime when $Ad\geq 1$. On the other hand, the minimum packing properties reached in our simulations are $\phi\approx 0.125$ and $Z\approx 2$ when $Ad$ is as high as 48 and $\mu_{\rm f}>1$, which is consistent with the conjectured asymptotic {\it adhesive loose packing} (ALP) limit \cite{Liu15}.

\begin{figure}
\begin{center}
\includegraphics[width=9cm]{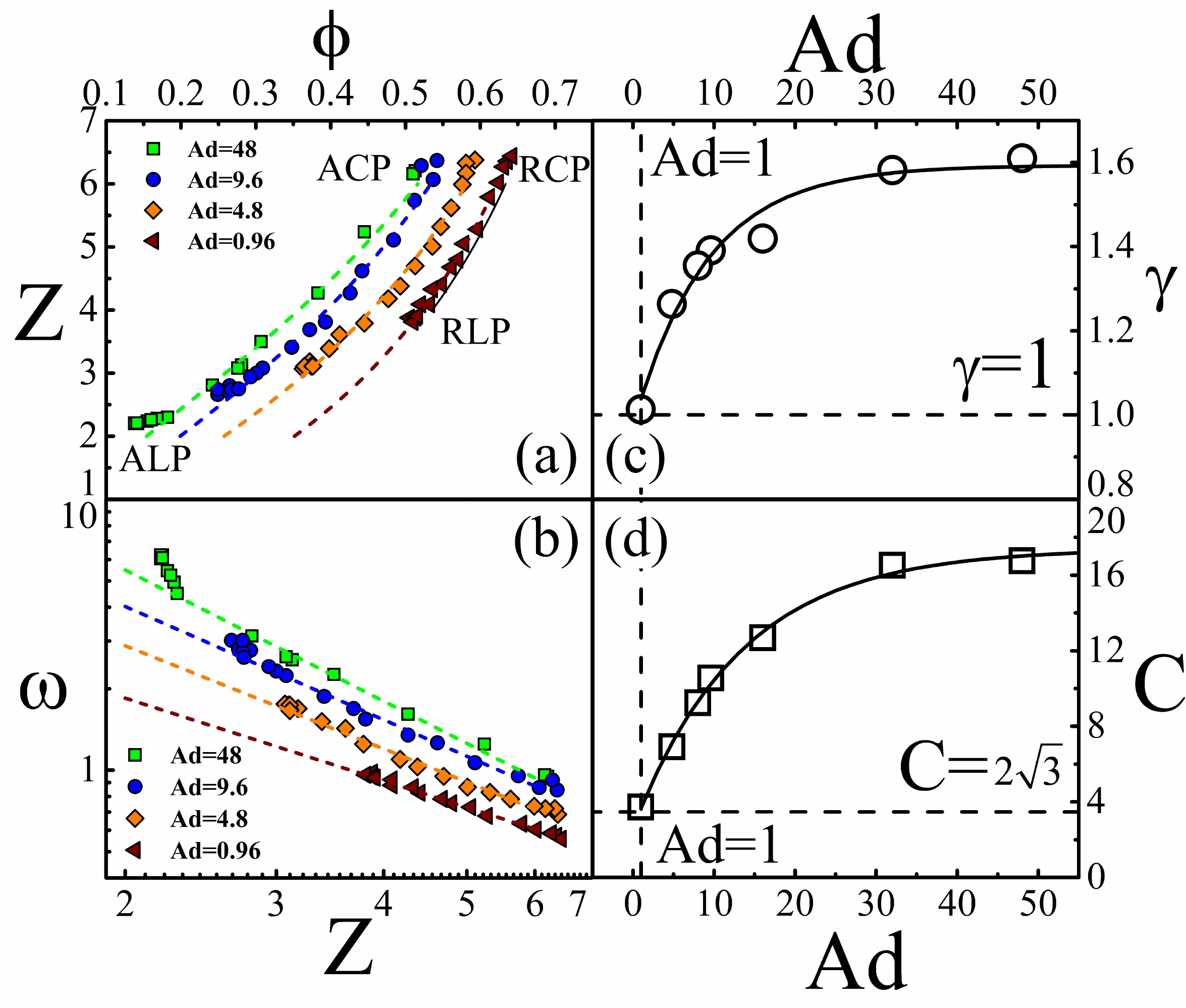}
\caption{\label{Fig:EOS} (a) Phase diagram of adhesive packings with changing $\mu_{\rm f}$ and $Ad$. The solid line is the theory line in \cite{Song08}. The dotted lines are the theoretical predictions Eq.(\ref{eq:EOS}). (b) The linear relation of $\omega$ and $Z$ on log-log coordinates. (c)(d) Variations of fitting parameters $C$ and $\gamma$ with $Ad$, respectively.}
\end{center}
\end{figure}

\begin{figure}
\begin{center}
\includegraphics[width=9cm]{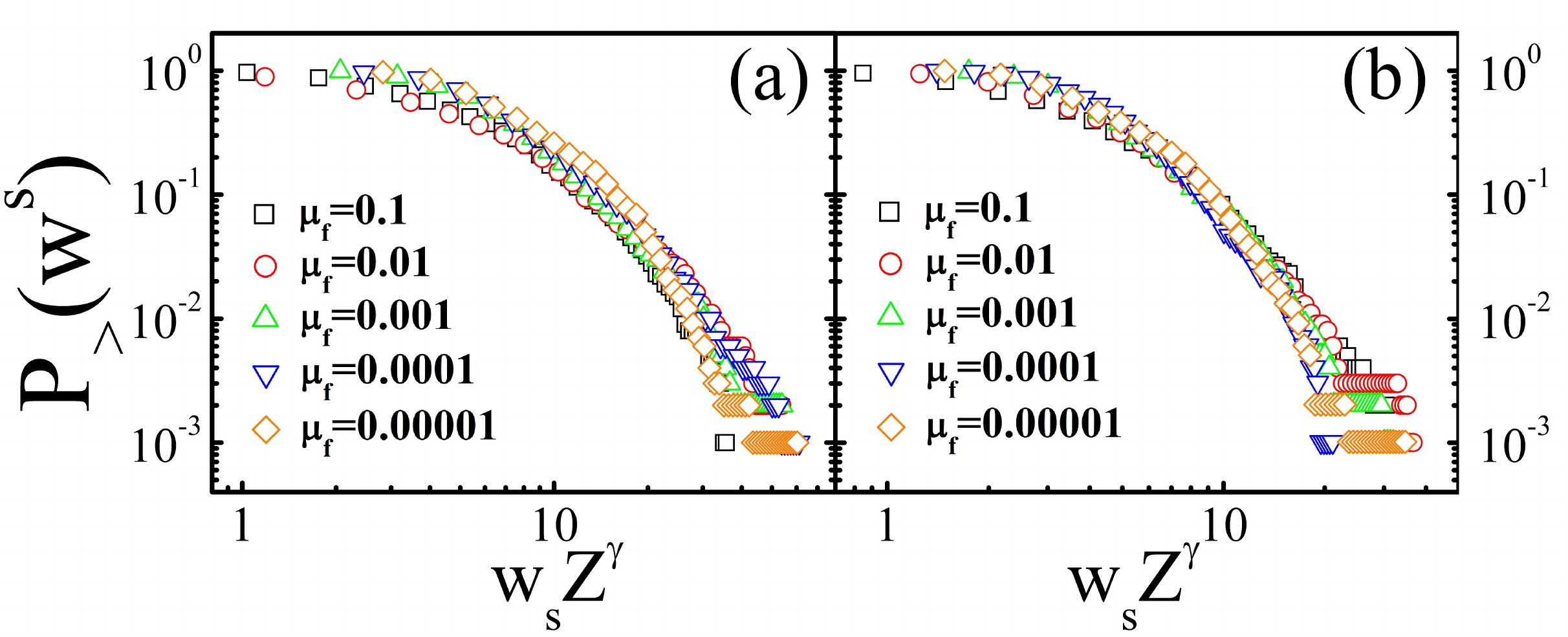}
\caption{\label{Fig:Pcumws} The inverse cumulative distribution function $P_>(\omega_s)$ as a function of $\omega_s Z^\gamma$. (a)(b) stand for cases of $Ad=48$ and $Ad=8$, respectively.}
\end{center}
\end{figure}

Next, we derive an analytical representation of the adhesive frictional equation of state (EOS) in the spirit of Edwards' ensemble approach at the mean-field level \cite{Song08,Jin10,Baule13}. We start with the Voronoi volume $W_i(\mu_{\rm f}, Ad)$ of a reference particle $i$ with a given $Ad$ and $\mu_{\rm f}$, which provides a tessellation of the total volume of the packing: $V=\sum_{i=1}^NW_i(\mu_{\rm f}, Ad)$. Here we define the average reduced free volume function $\omega$ for given $Ad$ and $\mu_{\rm f}$ as $\omega(\mu_{\rm f}, Ad)=\frac{\overline{W(\mu_{\rm f}, Ad)} - V_{\rm g}}{\overline{W(\mu_{\rm f}, Ad)}}$, where $V_{\rm g}$ is the volume of a single particle with radius $r_p$ in the packing, and $\overline{W(\mu_{\rm f}, Ad)}=\left<W_i(\mu_{\rm f}, Ad)\right>$ is the average volume of the Voronoi cell. This implies that $V=N \overline{W(\mu_{\rm f}, Ad)}$ and the packing fraction follows as $\phi(\mu_{\rm f}, Ad)=V_{\rm g}/\overline{W(\mu_{\rm f}, Ad)}=1/(\omega(\mu_{\rm f}, Ad)+1)$. The key step is to use a statistical mechanical description to obtain $\overline{W(\mu_{\rm f}, Ad)}$ or $\omega(\mu_{\rm f}, Ad)$, which can be calculated from the integral \cite{Song08}:
\be
\overline{W(\mu_{\rm f}, Ad)} = \int_{1/2}^\infty \ell^3 p(\ell;\mu_{\rm f}, Ad) d\ell.
\label{eq:W}
\ee 
Here $\ell$ is the distance from the center of a particle to its Voronoi boundary, and $p(\ell;\mu_{\rm f}, Ad)$ is the probability distribution function of $\ell$. By using the orientational reduced free volume function $\omega_s = 8 (\ell/r_p)^3 - 1$, we can rewrite Eq.(\ref{eq:W}) as:
\be
\omega(\mu_{\rm f}, Ad) = \int_{0}^\infty \omega_s p(\omega_s;\mu_{\rm f}, Ad) d\omega_s,
\label{eq:omega2}
\ee
where $p(\omega_s;\mu_{\rm f}, Ad)$ is the probability distribution function (PDF) of $\omega_s$;  $p(\omega_s; \mu_{\rm f}, Ad)=- \frac{P_>(\omega_s; \mu_{\rm f}, Ad)}{d\omega_s}$. Our computer simulations indicate that the inverse cumulative distribution function (CDF) $P_>(\omega_s;\mu_{\rm f}, Ad)$ has the form $P_>(\omega_s;\mu_{\rm f}, Ad)= q(Ad)f\left ( \omega_s [Z(\mu_{\rm f}, Ad)]^{\gamma(Ad)}; Ad \right)$, where the function $f(x; Ad)$ could be different for different $Ad$, and $Z$ depends on both $Ad$ and $\mu_{\rm f}$, and the parameters $q$ and the exponent $\gamma$ depend only on $Ad$. Figure~\ref{Fig:Pcumws} shows the inverse CDF $P_>(\omega_s)$ as a function of $\omega_s Z^\gamma$. The collapse of $P_>(\omega_s)$ of different $\mu_{\rm f}$ for a given $Ad$ reveals an invariant property of the microscopic structural PDF (see SI for details). Substituting $p_\omega(\omega_s)$ into Eq.(\ref{eq:omega2}), we thus obtain the equation of state as
\be
\omega(\mu_{\rm f}, Ad) = \frac{C(Ad)}{[Z(\mu_{\rm f}, Ad)]^{\gamma(Ad)} },
\label{eq:EOS}
\ee
where $C(Ad)=q(Ad)\left[-\int_0^\infty y f'(y; Ad) dy \right]$ and $y=\omega_s Z^{\gamma}$. The variations of parameters $C$ and $\gamma$ with $Ad$ are shown in Fig.\ref{Fig:EOS}c and \ref{Fig:EOS}d. In the non-adhesive case $Ad \to 0$, $\gamma(Ad \to 0) = 1$, $q(Ad \to 0) = 1$, and $f(y, Ad \to 0) = e^{-\frac{y}{2\sqrt{3}}}$. Plugging them into the expression of $C(Ad)$ we obtain $C(Ad\to 0) = 2\sqrt{3}$, which recovers the non-adhesive EOS, $\omega(\mu_{\rm f}, Ad \to 0)=\frac{2 \sqrt{3}}{Z(\mu_{\rm f}, Ad \to 0)}$ found in \cite{Song08}. Figure~\ref{Fig:EOS} shows the phase diagram of adhesive packings with changing $Ad$ and $\mu_{\rm f}$ (panel (a)) and the EOS (panel (b)). Accordingly, four different limit states can be identified as RCP, RLP, ALP and an adhesive close packing (ACP) state, where friction approaches zero but adhesion is very strong, as shown in Fig.~\ref{Fig:EOS}a. With the variation of $Ad$ and $\mu_{\rm f}$, most of the simulation results agree well with the theory Eq.(\ref{eq:EOS}) (Fig.\ref{Fig:EOS}a,\ref{Fig:EOS}b dashed lines), except for the deviation of some points with large $Ad(=48)$ and $\mu_{\rm f}(>0.1)$. We believe that this is due to the restriction of the ALP limit, since the packing properties cannot go below ALP as $Ad$ further increases.

\begin{figure}
\begin{center}
\includegraphics[width=7.5cm]{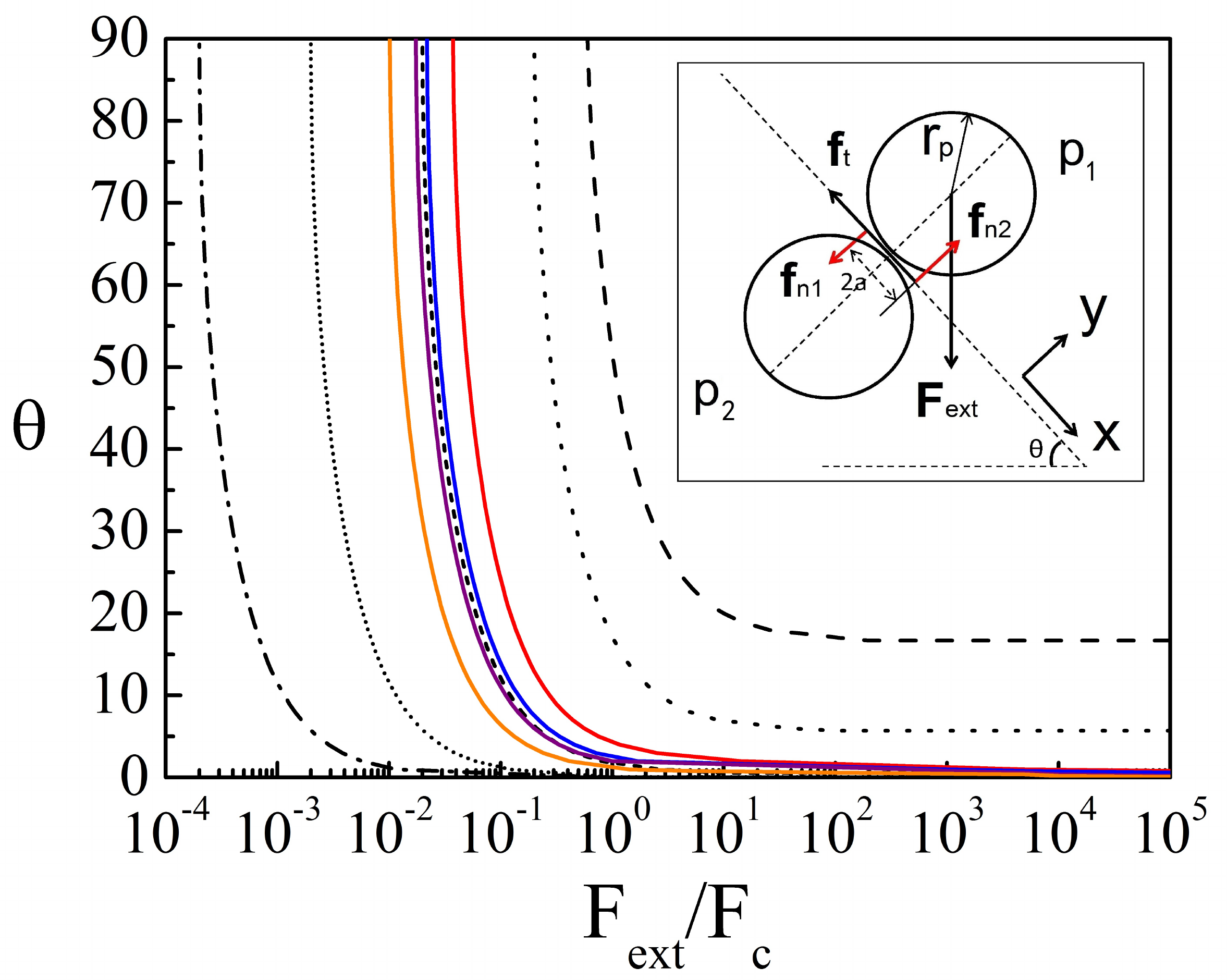}
\caption{\label{Fig:mecheqlmiu} Equilibrium diagram of two contact particles. The solid lines stand for the rolling equilibrium (re.) lines for $r_p=1,5,10,50\mu m$ from right to left, respectively. The others are sliding equilibrium (se.) lines with $\mu_{\rm f}=0.3,0.1,0.01,0.001,0.0001$ from right to left, respectively. The area under the lines indicates the equilibrium region that the particle will not roll or slide over another. The inset shows the schematic of force balance of two contact particles in 2D.}
\end{center}
\end{figure}

Compared with non-adhesive packings, a large number of particles with only one or two neighbours in adhesive packings can surprisingly be mechanically stabilized. Starting from the simplest case shown in the inset of Fig.\ref{Fig:mecheqlmiu}, we will explain how these particles reach mechanical equilibrium below. Without adhesion, the particle on the top will never be stabilized except for $\theta=0^{\circ}$. Nevertheless, in the presence of adhesion, when two contact particles start to roll or have the tendency of rolling, the rear side of the contact surface will be in touch until the critical pull-off force $F_C$ is reached, which is also known as the ``necking process''. As a consequence, the attractive normal stress on the rear side will provide additional rolling resistance to pull the particle not to roll over. Figure~\ref{Fig:mecheqlmiu} shows the equilibrium diagram produced with the parameters used in this work in terms of angle $\theta$ and external force $F_{ext}$ (see SI for detailed derivation). Note that the rolling equilibrium lines are only related to particle size but not sensitive. We can see that with very low external force, the adhesive particles can even be stabilized with $\theta=90^{\circ}$, which will never happen to non-adhesive granular matter. When $\mu_{\rm f} \geq 0.1$, the rolling equilibrium line lies under the sliding line, implying that particles roll first as they begin to have relative motion. In this case, the rearrangement during the packing formation is dominated by rolling, which agrees well with the fact that rolling is generally the preferred deformation mode for small adhesive particles \cite{Dominik97,Marshall14}. However, when $\mu_{\rm f} \leq 0.01$, sliding lines shift to the left of rolling line. In this case, particles will slide first and the additional rolling resistance caused by adhesion is not able to hold the moving particle, leading to prominent rearrangements of the packing. As can be observed from Fig.\ref{Fig:phiZ}, distinct increases of both $\phi$ and $Z$ occur when $\mu_{\rm f}<0.1$, which agrees with the above discussion.

For the adhesive loose packings with $2<Z<4$, it is obvious that the packing system is statically indeterminate in terms of the non-adhesive isostatic condition. Below we explore the modified counting arguments with both adhesion and friction on the basis of adhesive contact mechanics. Before that, let us recall the isostatic condition of non-adhesive packings \cite{Song08}. A packing is isostatic when the number of contact forces equals the number of force and torque balance equations: $N_n+N_t=E_f+E_t$, where $N_n, N_t, E_f, E_t$ are numbers of unknown normal forces, unknown tangential forces, force balance equations and torque balance equations, respectively. Thus, the average coordination number at the generalized isostatic point is derived as $Z(\mu_{\rm f})=2d\frac{1+1/2(d-1)f_1(\mu_{\rm f})}{1+(d-1)f_2(\mu_{\rm f})}$ (see details in Supplementary Information). Here $f_1(\mu_{\rm f})$ and $f_2(\mu_{\rm f})$ are two undetermined functions of friction coefficient that satisfy $f_i(0)=0$ and $f_i(\infty)=1$ for $i=1,2$. Then $Z(\mu_{\rm f})$ is reduced to the well-known $Z=2d$ for frictionless particles and $Z=d+1$ for infinitely rough ones. 


Here it is important to point out that for non-adhesive particles, typically of Hertz contact model, the real forces are symmetrically distributed in contact surface such that they can be equivalently simplified to point forces acting at the center with no additional torques. All the terms in the torque balance equations come from the point forces and there are no undetermined torque variables. However, for adhesive micron-sized particles, the phenomenon of material ``necking'' gives rise to an asymmetry of the forces in the contact region. As a result, more unknown torque variables will appear when simplifying the contact forces and the isostatic equation is modified as $N_n+N_t+N_T=E_f+E_t$, where $N_T$ is the number of the new unknown torques that can be expressed in 3d as $3/2NZg(\mu_{\rm f},Ad)$. $g(\mu_{\rm f},Ad)$ is a new undetermined function of $\mu_{\rm f}$ and $Ad$. It should be noted that here this function only works in dimension no more than 3, since it is derived based on the analysis of real forces. Similarly, the average coordination number is:
\be
\label{eq:ZmiuAd}
Z(\mu_{\rm f} ,Ad)=6\frac{1+f_1(\mu_{\rm f})}{1+2f_2(\mu_{\rm f})+3g(\mu_{\rm f} ,Ad)}.
\ee


The boundary conditions of the new function $g(\mu_{\rm f},Ad)$ are determined as follows: {\it (i)} If there is no friction, obviously there should be no additional torques regardless of whether there is adhesion and thus $g(\mu_{\rm f} =0,Ad)=0$. {\it (ii)} Then in case of no adhesion, Eq.(\ref{eq:ZmiuAd}) should go back to non-adhesive case, leading to $g(\mu_{\rm f} ,Ad=0)=0$. {\it (iii)} When friction and adhesion both go to infinity, all the particles should be constrained with additional torques and we have $g(\mu_{\rm f} =\infty,Ad=\infty)=1$. Substituting the above boundary conditions into Eq.(\ref{eq:ZmiuAd}), we have:
\be
\label{eq:ZmiuAdrs}
Z=\left\{
\begin{array}{cl}
6 &, \mu_{\rm f} = 0 \\
\frac{4}{1+g(Ad)} &, \mu_{\rm f} = \infty \\
2 &, \mu_{\rm f} = \infty, Ad=\infty
\end{array}
\right.
.
\ee
Further, we derive the function $g(\mu_{\rm f},Ad)$ as follows. From \cite{Wang10}, the function forms of $f_1(\mu_{\rm f})$ and $f_2(\mu_{\rm f})$ of non-adhesive packings are derived as $f_i(\mu_{\rm f})=\frac{k \mu_{\rm f}}{(1+k^2 \mu_{\rm f}^2)^{1/2}}$, where $k$ is the fitting parameter. Since $f_1(\mu_{\rm f})$ and $f_2(\mu_{\rm f})$ are not related to $Ad$, we can fix them and obtain the values of $g(\mu_{\rm f},Ad)$ for different $\mu_{\rm f}$ and $Ad$. Then the function form of $g(\mu_{\rm f},Ad)$ can be determined with a best fitting as, $g(\mu_{\rm f},Ad)=c_1(1-1/(1+a_1 Ad)^{b_1})(1-1/(1+a_2 \mu_{\rm f})^{b_2})$, where $a_1$, $b_1$, $c_1$, $a_2$ and $b_2$ are all fitting parameters (see SI for details). Substituting $f_1(\mu_{\rm f})$, $f_2(\mu_{\rm f})$ and $g(\mu_{\rm f},Ad)$ into Eq.(\ref{eq:ZmiuAd}) we get the prediction of $Z(\mu_{\rm f}, Ad)$. Together with the EOS in Eq.(\ref{eq:EOS}), $\phi(\mu_{\rm f}, Ad)$ can also be obtained. Both $\phi(\mu_{\rm f}, Ad)$ and $Z(\mu_{\rm f}, Ad)$ are also plotted in Fig.~\ref{Fig:phiZ} as dotted lines. We can see that the predictions are in good agreement with the simulations.


In summary, the influence of friction on random adhesive loose packings of uniform spherical micro-particles is examined by using adhesive contact dynamics simulations. The packing properties with different $\mu_{\rm f}$ and $Ad$ can be well described within a statistical ensemble theory framework. Most importantly, we fully expand the phase diagram of packings with arbitrary adhesion and friction \cite{Song08,Liu15}. Furthermore, we find that ALP is a well-defined limit of random adhesive packings in the perspectives of both dynamics and statics, and we identify a new limiting packing, ACP, for $\mu_f \to 0$ and $Ad \to \infty$. Parallel to RCP and RLP, experimental definitions for ALP and ACP, can be given as the minimum and maximum packing fraction one can get, when settling frictional, adhesive balls into a container without shaking \cite{Liu15}.

Acknowledgements. This work has been funded by the NNSFC (Nos. 51390491) and the NKBRPC (No. 2013CB228506). H. A. Makse acknowledges funding from NSF and DOE. We thank Profs. M. Doi, N. V. Brilliantov, J. S. Marshall for fruitful discussions on contact dynamic model, and Dr. G. Liu, M. Yang, H. Zhang, Mr. R. Tao, and W. Shi for their useful suggestions.

\bibliographystyle{apsrev4-1} 
\bibliography{refbib} 

\end{document}